\documentclass[letterpaper]{article}

\usepackage{times}
\usepackage{soul}
\usepackage{url}
\usepackage[utf8]{inputenc}
\usepackage{natbib}
\usepackage{amsmath}
\usepackage{amsthm}
\usepackage{amssymb}
\usepackage{amssymb}
\usepackage{booktabs}
\usepackage{algorithm}
\usepackage{algorithmic}
\usepackage{makecell}
\usepackage{pgfplots}
\pgfplotsset{compat=newest}
\usepackage[switch]{lineno}
\usepackage{color, colortbl}
\usepackage{subcaption}
\usepackage{booktabs, multirow}
\definecolor{clrLR}{RGB}{175, 213, 79}
\definecolor{clrXGB}{RGB}{32, 137, 227}
\definecolor{clMoTE}{RGB}{254, 111, 0}
\definecolor{clPAI}{RGB}{13, 6, 134}
\definecolor{clLG}{RGB}{134, 197, 189}
\definecolor{clOAI}{RGB}{129, 12, 134}
\definecolor{clAZ}{RGB}{252, 119, 147}

\usetikzlibrary{fit}
\newenvironment{customlegend}[1][]{%
\begingroup
\csname pgfplots@init@cleared@structures\endcsname
\pgfplotsset{#1}%
}{%
\csname pgfplots@createlegend\endcsname
\endgroup
}%
\def\addlegendimage{\csname pgfplots@addlegendimage\endcsname}

\usepackage{aaai24}  
\usepackage{times}  
\usepackage{helvet}  
\usepackage{courier}  
\usepackage{graphicx} 
\usepackage{natbib}  
\usepackage{caption} 
\usepackage{algorithm}
\usepackage{algorithmic}

\usepackage{newfloat}
\usepackage{listings}
\DeclareCaptionStyle{ruled}{labelfont=normalfont,labelsep=colon,strut=off} 
\floatstyle{ruled}
\newfloat{listing}{tb}{lst}{}
\floatname{listing}{Listing}
\title{MoJE: Mixture of Jailbreak Experts, Naive Tabular Classifiers as Guard for Prompt Attacks}
\author{
    Giandomenico Cornacchia\footnote{Corresponding author},
    Giulio Zizzo,
    Kieran Fraser,
    Muhammad Zaid Hameed,
    Ambrish~Rawat,
        Mark Purcell
}
\affiliations{
    IBM Research Europe\\
    Dublin, Ireland\\
    giandomenico.cornacchia1@ibm.com, giulio.zizzo2@ibm.com,
    kieran.fraser@ibm.com,
    zaid.hameed@ibm.com,
    ambrish.rawat@ie.ibm.com,
    markpurcell@ie.ibm.com
}

\usepackage{bibentry}

\begin{document}

\maketitle

\begin{abstract}
    The proliferation of Large Language Models (LLMs) in diverse applications underscores the pressing need for robust security measures to thwart potential jailbreak attacks. These attacks exploit vulnerabilities within LLMs, endanger data integrity and user privacy. Guardrails serve as crucial protective mechanisms against such threats, but existing models often fall short in terms of both detection accuracy, and computational efficiency. This paper advocates for the significance of jailbreak attack prevention on LLMs, and emphasises the role of input guardrails in safeguarding these models. We introduce MoJE (Mixture of Jailbreak Expert), a novel guardrail architecture designed to surpass current limitations in existing state-of-the-art guardrails. By employing simple linguistic statistical techniques, MoJE excels in detecting  jailbreak attacks while maintaining minimal computational overhead during model inference. Through rigorous experimentation, MoJE demonstrates superior performance capable of detecting 90\% of the attacks without compromising benign prompts, enhancing LLMs security against jailbreak attacks.
\end{abstract}

\section{Introduction}\label{sec:intro1}
The increasing adoption of Large Language Models (LLMs) in various applications brings forth the critical issue of ensuring their security against potential attacks, particularly jailbreak attacks. These attacks exploit vulnerabilities within the model and bypass existing safeguards to manipulate its behavior, posing significant risks to data integrity and user privacy. 
Thus, LLMs can be exploited by malicious actors for spreading misinformation, facilitating criminal activities ~\cite{Kreps_McCain_Brundage_2022,DBLP:journals/corr/abs-2301-04246, DBLP:journals/corr/abs-2302-05733} and even compromising scientific experimental settings ~\cite{Birhane2023}.
In response to this threat, the deployment of guardrails, which serve as protective mechanisms, has become essential to detect and mitigate such attacks~\cite{DBLP:conf/emnlp/WelblGUDMHAKCH21,DBLP:conf/emnlp/GehmanGSCS20, DBLP:journals/corr/abs-2402-01822}.

To prevent LLMs from producing undesired outputs~\cite{huang2023catastrophic}, several mitigation strategies have been proposed in the literature including training time and fine-tuning strategies e.g., instruction fine-tuning and Reinforcement Learning from Human Feedback (RLHF) ~\cite{carlini2024aligned}. These approaches have shown to improve the robustness of LLMs against jailbreak attacks, but often incur substantial computational costs and manual effort. On the other hand, guardrails-based mitigation approaches that can detect and filter the malicious input to the LLMs or post-process LLMs' output ~\cite{jain2023baseline,helbling2023llm,anwar2024foundational} provide a computationally efficient alternative for protecting LLMs against malicious attacks. Despite improving robustness against jailbreak attacks, these guardrail-based defence approaches are yet to prove effective against a more sophisticated attacker which remains an open research problem ~\cite{shayegani2023survey}.

To overcome the limitations of existing guardrail-based approaches against 
evolving attack strategies we propose a novel guardrail-based approach MoJE (Mixture of Jailbreak Expert), which outperforms current state-of-the-art guardrails in both attack detection accuracy, latency and throughput. Leveraging simple linguistic techniques, such as different tokenization strategies or n-gram feature extraction, MoJE demonstrates superior performance in identifying and neutralizing jailbreak attacks while maintaining minimal computational overhead during model inference. In addition, due to its modular nature, MoJE model can be easily extended to include models trained to defend against new attacks and out-of-distribution (OOD) datasets.

Our extensive experiments benchmark MoJE against both state-of-the-art open weight solutions, i.e., ProtectAI~\cite{deberta-v3-base-prompt-injection} and Llama-Guard~\cite{DBLP:journals/corr/abs-2312-06674}, and closed source  defences, i.e., OpenAI content moderation API~\cite{openaimoderation} and Azure AI Content Safety~\cite{azureaicontentsafety}. We demonstrate that MoJE not only surpasses these baselines in harmful content detection on various datasets but also exhibits superior resilience to jailbreak attacks. 


\section{Related Works}\label{sec:rel2}
The push for the safe and ethical deployment of advanced LLMs in digital contexts has spurred efforts in mitigating harmful content generation~\cite{burgess2023, doi/10.2860/333725}. The high volume of data used to train and poor data ingestion practice makes language models particularly vulnerable to prompt attacks. In the LLMs landscape, prompt attacks are generally known as ``\textit{jailbreak}''. Jailbreaks refers to the (un)intended manipulation of LLMs with linguistics and to force them into generating harmful or inappropriate content. 
Research efforts in defence against jailbreaks can be primarily divided into alignment-based and moderation-based approaches, each with distinct challenges and limitations~\cite{DBLP:journals/corr/abs-2402-01822, DBLP:journals/corr/abs-2403-13031}.

Alignment-based strategies, such as RLHF~\cite{DBLP:conf/nips/Ouyang0JAWMZASR22, DBLP:journals/corr/abs-2204-05862} and constitutional AI~\cite{DBLP:journals/corr/abs-2212-08073}, seek to align LLMs with ethical standards by training them to avoid engaging with predefined harmful topics. Despite progress, these methods require significant computational and human resources and mainly address pre-specified harmful content, limiting their effectiveness against new or evolving threats~\cite{DBLP:journals/corr/abs-2309-00614}. Moreover, fine-tuning often leads to surface-level modifications, as evidenced by persistently high logits of harmful tokens and vulnerability to subtle harmful behaviors~\cite{DBLP:journals/corr/abs-2401-05566}. Nevertheless, these methods face challenges from diverse disruptions such as new customization and manipulation techniques~\cite{DBLP:journals/corr/abs-2404-01833}. While jailbreak detection contributes to LLM security by identifying potential alignment breaches, it primarily identifies deviations rather than directly assessing harmfulness, inheriting the limitations of alignment-based approaches.
For the aforementioned reasons, new protective mechanisms to detect and mitigate such attacks are crucial~\cite{DBLP:conf/emnlp/WelblGUDMHAKCH21,DBLP:conf/emnlp/GehmanGSCS20, DBLP:journals/corr/abs-2402-01822}.

Moderation-based strategies, generally known as ``\textit{Guard}'' or ``\textit{Guardrails}'', initially aimed at improving social media safety, have shown promise in enhancing LLM safety. Traditional methods like the OpenAI Content Moderation API or Azure AI Content Safety API operate as classifiers trained on categorically labeled content~\cite{DBLP:conf/kdd/Lees0TSGMV22}. However, their effectiveness is limited to predefined categories, hindering their ability to address emerging risks or new attack typologies. Recent approaches also employ general pre-trained LLMs, e.g., Llama-Guard \cite{DBLP:journals/corr/abs-2312-06674}  (a finetuned version of Llama2 which benefits from its broader contextual understanding) for more extensive harmful content detection. However, these methods also inherit vulnerabilities from associated to their base LLMs, particularly susceptibility to sophisticated jailbreak attacks
or misalignment after fine-tuning~\cite{DBLP:journals/corr/abs-2310-03693}. Furthermore, these models require high computational effort as they are generally in the order of millions or billions of parameters. Thus, the use of such LLM-based guardrails in real-time applications necessitates a powerful GPU-ready cloud architecture, which inevitably leads to growth in latency, energy costs, and the application's overall carbon footprint.

In this context, MoJE extends the moderation-based approach, with a simple and robust moderation framework resistant to adversarial attacks with low computational requirements (i.e., a single CPU-core can predict with a low latency).


\section{Methodology}\label{sec:meth3}
\subsection{Preliminaries}
\begin{figure*}[!htbp]
    \centering
    \begin{subfigure}{0.49\textwidth}
        \includegraphics[width=\linewidth]{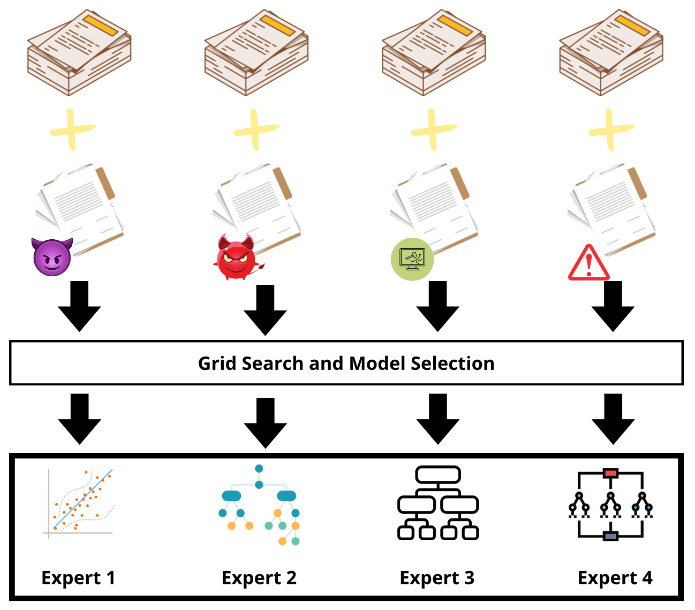}
        \caption{Training pipeline of MoJE}
        \label{fig:subfiga}
    \end{subfigure}
    \begin{subfigure}{0.49\textwidth}
        \includegraphics[width=\linewidth]{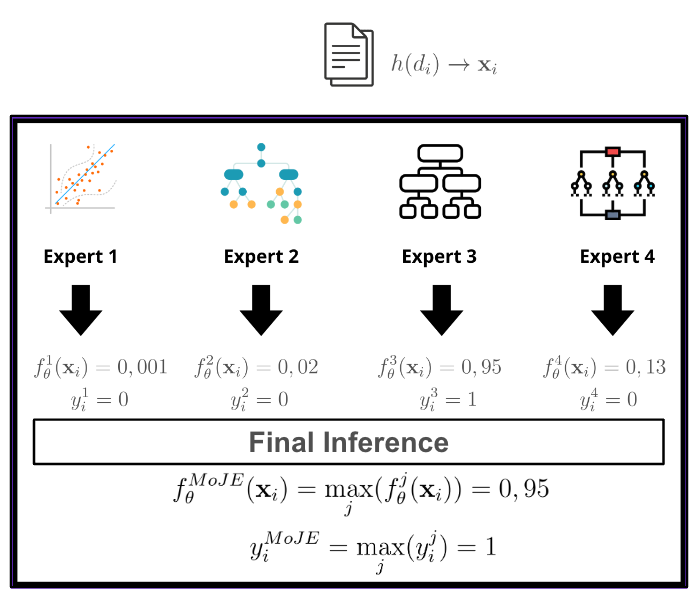}
        \caption{Inference pipeline of MoJE}
        \label{fig:subfigb}
    \end{subfigure}\caption{Figure~\ref{fig:subfiga} shows the training phase where a model is trained for each jailbreak dataset. This employs both a grid search over model hyperparameters and a model selection process. This trained model is denoted as the $f_{\theta}^j$ expert in the Mixture of Jailbreak Experts. Figure~\ref{fig:subfigb} represents the inference phase. If the max posterior probability of MoJE is higher than a set threshold $\tau$, we take the maximum prediction probability. Otherwise, framework averages the prediction probability of all the expert models.
}
    \label{fig:twosubfig}
\end{figure*}

Let $\mathcal{D} = \{d_i,y_i\}_{i=1}^N$ be a dataset of $N$ documents, or prompts, $d_i$ and associate label $y_i$, where $y_i \in \mathrm{Y} = \{jailbreak, \ benign\}$. Furthermore, each sample $d_i$ with $y_i=\{jailbreak\}$ can belong to a specific category of jailbreak. Thus, the dataset $\mathcal{D}$ can be divided into $l+1$ different subsets which $\mathcal{D}|_{benign}$ is the set of \textit{benign} documents while $\mathcal{D}|_{jailbreak}^j$, with $j$ from 1 to $l$ and $y_i=jailbreak$, the $j$-th specific set of jailbreaks.

We assume that our dataset, after appropriate transformation, can be represented as a $m$ dimensional random variable $\mathrm{X} \in \mathbb{R}^m$ through a function $h:\mathcal{D}\rightarrow\mathrm{X}$, without considering the associate label $y_i$ in the transformation of each document $d_i$.
Since the definition domain of $\mathrm{Y}$ is a binary set, we can define the problem as a binary classification task and learning a mapping function $f_{\Theta}^{MoJE}:\mathrm{X}\rightarrow\mathrm{Y}$ is our main goal.~\footnote{It could be argued that this task is suited for multi-classification, but due to linguistic complexities and overlapping (relatively new) concepts between jailbreak categories, binary classification was selected for simplicity. Multi-classification is remanded for future investigation.} 

\subsection{Mixture of Jailbreak Experts}\label{sec:MoJE}

Let's consider having $l$ classifiers, where each classifier is trained on $\mathcal{D}|_{b\cup j} = \mathcal{D}|_{benign}\cup\mathcal{D}|_{jailbreak}^j$, i.e., the set of benign prompts and the $j$-th specific jailbreak prompt which is transformed into a vector space $\mathrm{X}_j$ after applying the $h(\cdot)$ transformation function.
Thus, it can be considered the $j$-th jailbreak expert.
To this end, we find each model parameters $\hat{\theta}_j$ that maximizes the posterior probability over the respective $j$-th training set, i.e.:

\begin{equation}
\hat{\theta}_j = \underset{\theta}{\arg \max } \,\mathrm{P}(\theta|\mathrm{X}_j).
\end{equation}

Our MoJE will be the ensemble of each model parameters as $\Theta=\{\hat{\theta}_1, \hat{\theta}_2, \dots, \hat{\theta}_l\}$.
 For the sake of simplicity, the predictions of each classifier is denoted as $f_{\theta}^j(\cdot)$ which represent the probability assigned by the $j$-th classifier to the positive class (i.e., identified as the $j$-th jailbreak) and as $f_{\Theta}^{MoJE}(\mathbf{x}_i)$ the probability assigned by our MoJE to the positive class. 
Considering a prompt sample $d_i$ that is transformed in a vector $\mathbf{x}_i$ i.e., $\mathbf{x}_i = h(d_i)$, the inference of our final MoJE classifier is defined as:

\begin{equation}
  f_{\Theta}^{MoJE}(\mathbf{x}_i)=
    \begin{cases}
      \max \mathcal{P}(\mathbf{x}_i) & \text{if } \max \mathcal{P}(\mathbf{x}_i) >= \tau\\
      \text{avg} \mathcal{P}(\mathbf{x}_i) & \text{if } \max \mathcal{P}(\mathbf{x}_i) < \tau\\
    \end{cases}       
\end{equation}

where $\mathcal{P}(\mathbf{x}_i) = \{f_{\theta}^1(\mathbf{x}_i), f_{\theta}^2(\mathbf{x}_i), \ldots, f_{\theta}^l(\mathbf{x}_i)\} \in [0, 1]^l$ denotes a vector of all classifiers' posterior probability, and $\tau$ denotes the probability threshold which is set to $\tau = 0.5$ for mathematical convenience.

In conclusion, our \textit{Mixture of Jailbreak Expert} is an ensemble of tabular classifiers that selects the maximum prediction probability value if one or more of the classifiers' posterior probability is equal or greater than the threshold $\tau$ and the average in the case all predictions probabilities are below that threshold. The average of all classifiers' prediction probability represents a classic ensemble vote prediction in which the uncertainty of the posterior probability is averaged across all the classifiers. Figure~\ref{fig:twosubfig} represents the model training phase and inference phase for our proposed approach MoJE.

\section{Experimental Setting}\label{sec:expset4}
This section describes the dataset, SOTA guardrails, models, and tuning strategy used in this work. 
\begin{table}[!t]\centering
\begin{tabular}{lrrrr}\toprule
Dataset &$|\mathcal{D}|_i$ &prompt type & $\mathcal{Y}$\\\midrule
harmful behaviors &512 &harmful ques. &\textit{jailbreak} \\
gandalf &1000 &instruction &\textit{jailbreak} \\
gcg-vicuna &512 &suffix style &\textit{jailbreak} \\
jailbreak prompts &666 &instruction &\textit{jailbreak} \\\midrule
puffin &6994 &user-convers. &\textit{benign} \\
alpaca &52002 &instruction &\textit{benign} \\
awes. chatgpt p. &152 &instruction &\textit{benign} \\
\bottomrule
\end{tabular}

\caption{List of datasets characteristics used to train our models and evaluate the same against the competitors. Each dataset can be devided into \textit{jailbreak} and \textit{benign} category as target class.}\label{tab:datasetStats}
\end{table}

\subsection{Dataset}
We will divide this section as follows: \textit{jailbreak-prompts dataset}, in which we list the jailbreak dataset we used for our analysis, and \textit{benign-prompt dataset}, in which we list the employed benign dataset. The use of a benign dataset is crucial to prevent the detector being biased toward jailbreaks with the consequence of a high false positive rate (FPr) which is not desirable in deployment scenarios. Datasets are listed in Table~\ref{tab:datasetStats} along with their characteristics.

\subsubsection{Jailbreak-Prompts Dataset}
Here, we present the datasets used for jailbreak prompts, i.e., \textit{harmful behaviors}, \textit{gandalf ignore instructions}, \textit{gcg-vicuna}, and \textit{jailbreak prompts}.\\
\noindent\textbf{Harmful Behaviors:}
The Harmful Behaviors dataset is a sub-set of the AdvBench dataset, which has been designed to test LLM alignment for safety requirements~\cite{DBLP:journals/corr/abs-2307-15043}. It is divided in two sub-sets: Harmful Strings and Harmful Behaviors. Both datasets have been generated by prompting {\tt Wizard-Vicuna-30B-Uncensored}, an uncensored and unaligned version of Vicuna model. The authors handcrafted 100 and 50 prompts respectively. Then, they generated 10 new samples, each giving a 5-shot demonstration as prompt. The latter was selected for this work as they are more aligned with the jailbreak setting.  
The dataset consists of 512 harmful behavior prompts crafted as instructions, encompassing the same themes within the harmful strings setting. The prompts are crafted as questions as the aim is to enforce the model to generate harmful content as a response that complies with the harmful instructions. It spans various themes observed in online interactions, including cyberbullying, hate speech, and harassment, thereby serving as a crucial resource for training and assessing algorithms tasked with detecting and curtailing harmful behavior within digital environments and online communities.

\noindent\textbf{Gandalf Ignore Instruction:}
The Gandalf Ignore Instruction dataset\footnote{\url{https://huggingface.co/datasets/Lakera/gandalf_ignore_instructions}} consists of prompts collected by Lakera AI~\cite{gandalf_ignore_instructions}. The prompts were collected during an educational game designed to inform people about AI leakage risks of prompt attacks on large language models (LLMs). 
The dataset consists of 1000 instruction-based prompts that utilize role-playing to circumvent the model's alignment defense such that it reveal the game's secret password.

\noindent\textbf{GCG Vicuna behavior:}
The GCG-Vicuna dataset was generated using the technique and methodology outlined by~\cite{DBLP:journals/corr/abs-2307-15043}. This dataset consists of 512 samples - the same number found in the Harmful Behaviors dataset, as it was used for prompting Vicuna model during the attack. The type of attack performed is the ``individual'' GCG (Greedy Coordinate Gradient) method. For each harmful behavior prompt, the attack starts by appending an adversarial suffix of twenty spaced exclamation marks (i.e., ``! '') to the prompt. The attack subsequently makes further revisions to the appended suffix, attempting to reduce the loss, until the model deigns to answer without refusal keywords. Throughout the attack, several distinct attack suffixes may be generated. The suffix selected was one which resulted in a successful attack and has lowest loss. To generate and test the performance of the new suffix, we employed the {\tt Vicuna-7b-v1.5}\footnote{\url{https://huggingface.co/lmsys/vicuna-7b-v1.5}} model~\cite{DBLP:conf/nips/ZhengC00WZL0LXZ23}, a fine-tuned version of Llama2~\cite{DBLP:journals/corr/abs-2307-09288}, replicating the experimental setup of \cite{DBLP:journals/corr/abs-2308-14132}.

\noindent\textbf{Jailbreak Prompts:}
The Jailbreak Prompts dataset comprises examples of four platforms (i.e., Reddit, Discord, websites, and open-sources datasets) from December 2022 to May 2023, which consists of 6387 prompts, then filtered to 666 prompt considered as jailbreaks ``in the wild" by \cite{DBLP:journals/corr/abs-2308-03825}.

\subsubsection{Benign Prompts}
The following datasets used in this work are defined as benign prompts: \textit{puffin}, \textit{alpaca},  and \textit{awesome chatgpt prompt}.\\

\noindent\textbf{Puffin:}
The Puffin dataset is a collection of multi-turn conversations between GPT-4 and humans\footnote{\url{https://huggingface.co/datasets/LDJnr/Puffin}}. This dataset is comprised of 2000 conversations, with an average of 10 turns, and contains conversation context lengths stretching over 1000 tokens. In the context of this work, only the set of 6,994 prompts produced by the human side of the conversation was selected, as these prompts align best with benign labeled data. 

\noindent\textbf{Alpaca:}
The Alpaca dataset\footnote{https://huggingface.co/datasets/tatsu-lab/alpaca}, comprises 52,000 instructions and demonstrations generated by OpenAI's {\tt text-davinci-003} engine~\cite{alpaca}. The primary use case of the Alpaca dataset is to serve as a valuable resource for instruction-tuning language models, facilitating enhanced adherence to instructions. Indeed, the data generation pipeline was explicitly tasked for instruction data generation. In our setting, the distilled instruction will consist of benign instruction prompts that contrast the effect of role-playing jailbreak instruction prompts.

\noindent\textbf{Awesome ChatGPT Prompt:}
Awesome ChatGPT Prompts is a repository containing a curated collection of prompt examples designed to be used with the ChatGPT model\footnote{\url{https://huggingface.co/datasets/fka/awesome-chatgpt-prompts}}. Specifically, it offers a prompt collection designed to be successful on use cases applied to ChatGPT models. For our use case, it increases the collection of prompts associated with a benign role-playing scenario.

\subsection{Guardrails}
In this section we introduce the set of SOTA Guardrails or chat Moderation tools that we use for comparison with our proposed approach.

\noindent\textbf{ProtectAI:}
ProtectAI guard\footnote{\url{https://huggingface.co/protectai/deberta-v3-base-prompt-injection}} is a security tool designed to identify and prevent prompt injection attacks which can manipulate language models into producing unintended outputs~\cite{deberta-v3-base-prompt-injection}. 

The model is a fine-tuned version of the {\tt microsoft/deberta-v3-base}\footnote{\url{https://huggingface.co/microsoft/deberta-v3-base}} model, based on the Microsoft BERT Language Model with 86 million backbone parameters~\cite{he2021deberta}. The ProtectAI guard is trained on a combination of prompt injections, jailbreak, and benign prompt datasets. It categorizes inputs into two classes: 0 for no injection and 1 for detected injection.

\noindent\textbf{Llama-Guard:}
Llama-Guard\,\footnote{\url{https://huggingface.co/meta-llama/LlamaGuard-7b}} is an LLM-based safeguard model tailored for Human-AI conversation scenarios~\cite{DBLP:journals/corr/abs-2312-06674}. The model is based on the family of open weight Llama2 models deployed by Meta~\cite{DBLP:journals/corr/abs-2307-09288}. Indeed, it is an instruction based tuned version of {\tt Llama2-7b}. It employs a curated safety risk taxonomy template to effectively categorize prompts and responses, enhancing safety assessment and moderation. This model acts as a binary classification model with the first generated token, i.e., ``\textit{safe}'' or ``\textit{unsafe}''), categorizing the prompt. If the model assessment is ``\textit{unsafe}'', then the model generates a new line, listing the taxonomy categories that are violated in the given piece of content. Note that, as a text-to-text approach, it does not contain the output probability, hence the AUC result is not present in Table~\ref{tab:aggregateResults} for Llama-Guard model.

\noindent\textbf{OpenAI Moderator API:}
OpenAI Moderator\footnote{\url{https://platform.openai.com/docs/guides/moderation/overview?lang=python}} is an API AI-powered content moderation system designed to check, monitor, and filter user-generated content that is potentially harmful~\cite{openaimoderation}. Leveraging natural language processing (NLP) algorithms and machine learning models, OpenAI Moderator automatically identifies and flags potentially harmful or inappropriate content, such as hate speech, spam, abusive language, harmful intent and instructions. The model engine used at the time of experiments is the {\tt text-moderation-007}. The output of the model is divided into 11 categories which can be flagged separately. Each of these categories is linked with a probability. We treat the problem as a binary classification task, where the probability of a jailbreak is the  maximum of the harmful categories in cases where more than one category is flagged.

\noindent\textbf{Azure AI Content Safety API:}
Azure AI Content Safety API\footnote{\url{https://learn.microsoft.com/en-us/azure/ai-services/content-safety/}} is a cloud-based service offered by Microsoft Azure for content moderation and safety analysis~\cite{azureaicontentsafety}. It consist of an ensemble of classification models to identify and prevent the output of harmful content. This system actively detects and addresses specific categories of potentially harmful content within both input prompts and output completions, with dedicated models trained and tested for hate speech, sexual content, violence, and self-harm across multiple languages. While the service extends its support to various languages beyond the specified ones, users are advised to conduct their own testing to verify its effectiveness for their specific application needs. We specifically employed the jailbreak endpoint API\footnote{The version used for the current experiment refers to the {\tt 2023-10-01-preview}}. In the same manner as with Llama Guard, we do not obtain a probability but only a \textit{boolean} jailbreak flag.

\subsection{Data Preprocessing}
To keep our proposed approach computationally efficient, we employed simple classifiers to perform the detection. To comply with classic tabular classifier requirements, we transform the prompt input or document i.e., $d_i$ into a vector $\mathbf{x}_i$ through a transform function $h(\cdot)$. We used the $n$-gram occurrences count\footnote{\url{https://scikit-learn.org/stable/modules/generated/sklearn.feature_extraction.text.CountVectorizer.html}}, with $n=1$ (i.e., uni-gram), as the feature extraction function to map the dataset into an $m$-dimensional vector (i.e., $h(\mathcal{D})=\mathrm{X}$). This  dimension `$m$' of the features' vector space $\mathrm{X}:\mathbb{R}^m$ depend on two factors: (i) the number of tokens detected based on the splitting strategy and (ii) the feature extraction strategy (e.g., uni-gram vs bi-gram). Furthermore, we separate punctuation from the words and treat it as a separate token since few attacks are based on special punctuation tokens as a prefix or suffix (e.g., \textit{gcg-based attack}).

We used two models: a simple Logistic Regression\footnote{\url{https://scikit-learn.org/stable/modules/generated/sklearn.linear_model.LogisticRegression.html}} (LR) model, for understanding the linear dependencies between each attack type, and eXtreme Gradient Boost Machine\footnote{\url{https://xgboost.readthedocs.io/en/stable/}} (XGB), as one of the most suitable and popular classifiers used for tabular datasets. For our proposed MoJE framework, the $j$-th expert in the model selection phase is chosen from the two models (LR and XGB) that are obtained after a grid-search on model hyperparameters  (see Figure~\ref{fig:subfiga}). 

Furthermore, we also use LR and XGB as baseline models trained on the full dataset $\mathcal{D}$ to compare the effectiveness of our proposed MoJE framework.

\subsection{Tuning Strategy}
The datasets have been divided using the hold-out method, with an initial 80/20 train-test split, and a further splitting of the train dataset into an 80/20 train-validation split. The validation set facilitates hyperparameter tuning and expert model selection, whilst the test set facilitates identifying the best model in the Results' discussion. Finally, we divide the training datasets in chunks of $j$-th jailbreak and benign prompts. For each of these chunks we tune each classifier $f_{\theta}^j$ with a 5-fold cross-validation grid search methodology where for each set of parameters we choose the set that maximize the $F_{\beta}$ score, with $\beta=0.5$, defined as:   

\begin{equation}
F_\beta=\left(1+\beta^2\right) \cdot \frac{\text { precision } \cdot \text { recall }}{\left(\beta^2 \cdot \text { precision }\right)+\text { recall }}.
\end{equation}

We decided to optimize the models with $F_{\beta}$ score to reduce the FP rate while still maintaining a relatively high TP rate instead of balancing TP rate and TN rate. We consider this setting because we deem FP rate to be more important to the usefulness of the detector in a real application scenario.

After training each model on a $j$-th dataset subset, we choose the best model and call it as a $j$-th jailbreak expert, based on the best $F_\beta$, with $\beta = 0.5$. Thus, we ensemble all the best $j$-th dataset subset models for each $j$ to create a MoJE pipeline as described in Figure~\ref{fig:twosubfig} and Section~\ref{sec:MoJE}.

The advantage of our approach is the modularity. Indeed, if a new type of attack is discovered, a new classifier can be easily trained and added to the mix (i.e., $f_{\theta}^{l+1}$). On the other hand, for Llama-Guard or ProtectAI, it is necessary to retrain the whole model.

For a baseline defence comparison, we also trained LR and XGB on the whole dataset $\mathcal{D}$ respectively to account for a simpler approach for comparison to our proposed MoJE.

\section{Results}\label{sec:res5}
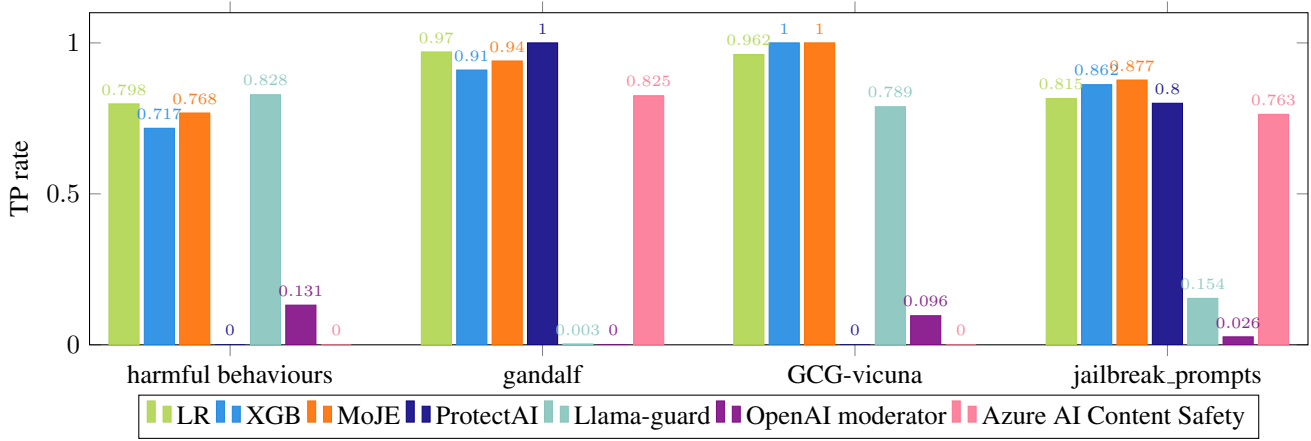
\begin{figure*}[!ht]
    \centering
    \pgfplotsset{width=\textwidth,compat=1.17}
    \begin{tikzpicture}
        \begin{axis}[
            ybar,
            ylabel={TP rate},
            height=6cm,
            ymin=0,
            ymax=1.1,
            xtick=data,
            xticklabels={harmful behaviours, gandalf, GCG-vicuna, jailbreak\_prompts},
            yticklabel={\pgfmathprintnumber\tick},
            bar width=0.4cm,
            nodes near coords={\tiny\pgfmathprintnumber[fixed,precision=3]{\pgfplotspointmeta}},
            nodes near coords align={vertical},
            enlarge x limits=0.15,
            legend style={at={(0.5,-0.15)},
                anchor=north,legend columns=-1},
        ]
        
        \addplot[ybar, fill=clrLR, fill opacity=0.9, mark size=3pt, color=clrLR] coordinates {(1, 0.7980) (2, 0.9700) (3, 0.9615) (4, 0.8154)};
        \addplot[fill=clrXGB,fill opacity=0.9, mark size=3pt, color=clrXGB,
    ] coordinates {(1, 0.7172) (2, 0.9100) (3, 1.0000) (4, 0.8615)};
        \addplot[fill=clMoTE,fill opacity=0.9, mark size=3pt, color=clMoTE,
    ] coordinates {(1, 0.7677) (2, 0.9400) (3, 1.0000) (4, 0.8769)};
        \addplot[fill=clPAI,fill opacity=0.9, mark size=3pt, color=clPAI,
    ] coordinates {(1, 0.0000) (2, 1.0000) (3, 0.0000) (4, 0.8000)};
        \addplot[fill=clLG,fill opacity=0.9, mark size=3pt, color=clLG,
    ] coordinates {(1, 0.8283) (2, 0.0028) (3, 0.7885) (4, 0.1538)};
        \addplot[fill=clOAI,fill opacity=0.9, mark size=3pt, color=clOAI,
    ] coordinates {(1, 0.1313) (2, 0.0000) (3, 0.0962) (4, 0.0263)};
        \addplot[fill=clAZ,fill opacity=0.9, mark size=3pt, color=clAZ,
    ] coordinates {(1, 0.0000) (2, 0.8250) (3, 0.0000) (4, 0.7632)};
        
        \legend{LR, XGB, MoJE, ProtectAI, Llama-guard, OpenAI moderator, Azure AI Content Safety}
        \end{axis}
    \end{tikzpicture}
    \caption{TP rate for each jailbreak dataset (i.e., \textit{harmful behaviors}, \textit{gandalf ignore instructions}, \textit{gcg-vicuna}, and \textit{jailbreak prompts}) given our tabular models (i.e., LR, XGB, and MoJE), the open-weight models (i.e., ProtectAI and Llama-Guard), and closed source one (i.e., OpenAI moderator and Azure AI Content Safety).}
    \label{fig:histogramTP}
\end{figure*}

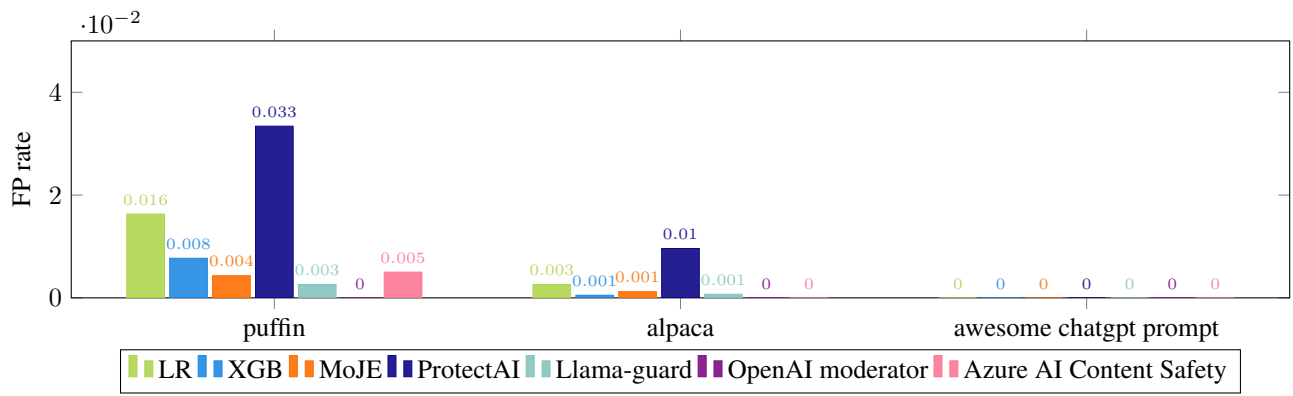
\begin{figure*}[htbp]
    \centering
    \pgfplotsset{width=\textwidth,compat=1.17}
    \begin{tikzpicture}
        \begin{axis}[
            ybar,
            xlabel={Dataset},
            ylabel={FP rate},
            height=5cm,
            ymin=0,
            ymax=0.05,
            xtick=data,
            xticklabels={puffin, alpaca, awesome chatgpt prompt},
            yticklabel={\pgfmathprintnumber\tick},
            bar width=0.5cm,
            nodes near coords={\tiny\pgfmathprintnumber[fixed,precision=3]{\pgfplotspointmeta}},
            nodes near coords align={vertical},
            enlarge x limits=0.25,
            legend style={at={(0.5,-0.20)},
                anchor=north,legend columns=-1},
        ]
        
        \addplot[ybar, fill=clrLR, fill opacity=0.9, mark size=3pt, color=clrLR] coordinates {(1, 0.0163) (2, 0.0026) (3, 0.0000)};
        \addplot[fill=clrXGB,fill opacity=0.9, mark size=3pt, color=clrXGB,
    ] coordinates {(1, 0.0077) (2, 0.0005) (3, 0.0000)};
        \addplot[fill=clMoTE,fill opacity=0.9, mark size=3pt, color=clMoTE,
    ] coordinates {(1, 0.0043) (2, 0.0012) (3, 0.0000)};
        \addplot[fill=clPAI,fill opacity=0.9, mark size=3pt, color=clPAI,
    ] coordinates {(1, 0.0334) (2, 0.0096) (3, 0.0000)};
        \addplot[fill=clLG,fill opacity=0.9, mark size=3pt, color=clLG,
    ] coordinates {(1, 0.0026) (2, 0.0007) (3, 0.0000)};
        \addplot[fill=clOAI,fill opacity=0.9, mark size=3pt, color=clOAI,
    ] coordinates {(1, 0.0000) (2, 0.0000) (3, 0.0000)};
        \addplot[fill=clAZ,fill opacity=0.9, mark size=3pt, color=clAZ,
    ] coordinates {(1, 0.0050) (2, 0.0000) (3, 0.0000)};
        
        \legend{LR, XGB, MoJE, ProtectAI, Llama-guard, OpenAI moderator, Azure AI Content Safety}
        \end{axis}
    \end{tikzpicture}
    \caption{FP rate for each benign dataset (i.e., \textit{puffin}, \textit{alpaca}, and \textit{awesome chatgpt prompt}) given our tabular models (i.e., LR, XGB, and MoJE), the open-weight models (i.e., ProtectAI and Llama-Guard), and closed source one (i.e., OpenAI moderator and Azure AI Content Safety).}
    \label{fig:histogramFP}
\end{figure*}

In this section, we will present and discuss the results of our investigation. 
We first present the aggregate results of all the datasets, benign and jailbreak, on the test set split. Then, we have a more extenisve analysis of jailbreak detection (i.e.,  TP rate) and benign miss-classification (i.e., FP rate) results presented respectively in Figure~\ref{fig:histogramTP} and Figure~\ref{fig:histogramFP}.


\begin{table}[!htp]\centering
\small
\begin{tabular}{lccrccr}\toprule
$f_{\theta}(X)$ &AUC & ACC &$\text{F}_{\beta=0.5}$ &Recall &Precis. \\\midrule
LR &0.9816 &0.9918 &0.9096  &0.8987 &0.9124 \\
XGB &0.9946 &0.9936 &0.9513  &0.8799 &0.9710 \\
MoJE &\textbf{0.9947} &\textbf{0.9944} & \textbf{0,9529}  &\textbf{0.9043} &0.9659 \\\midrule
ProtectAI &0.7878 &0.9697 &0.6594  &0.5704 &0.6862 \\
Llama-G. &- &0.9711 &0.7159  &0.3602 &0.9505\\\midrule
openAI m. &0.8446 &0.5212 &0.2235  &0.0544 &\textbf{1.0000} \\
AzureAPI &- &0.7153 & 0.7944 &0.4399 &0.9949 \\
\bottomrule
\end{tabular}\caption{Classification results of each model on AUC, ACC, $F_{\beta}$ with $\beta=0.5$, Recall, and Precision. The model are divided into three section: (i) model that belongs to our pipeline, (ii) open-source fine tuned Language Models, and (iii) closed-source API endpoint. Best results are in bold.}\label{tab:aggregateResults}
\end{table}

We start by presenting the aggregate results of each model with respect to the full test set. Results are condensed in Table~\ref{tab:aggregateResults} which presents the classification results of various models across multiple metrics, including AUC, accuracy (ACC), $F_\beta$ with $\beta = 0.5$, recall, and precision. The models are categorized into three sections: (i) models belonging to our pipeline (i.e., naive tabular classifiers), (ii) open-weights fine-tuned Language Models, and (iii) closed-source API endpoints. Among the models in our pipeline, MoJE demonstrates the highest performance across most metrics, achieving an AUC of 0.9947 and an ACC of 0.9944. Notably, MoJE also attains the highest $F_\beta$ score of 0.9529, indicating strong performance in balancing precision and recall. However, ProtectAI and LLAMA-guard, while achieving great ACC scores, show comparatively lower performance in terms of $F_\beta$ score, demonstrating potential areas of weakness in their predictive capabilities.

On the other hand, models sourced from open-source fine-tuned Language Models and closed-source API endpoints exhibit a more diverse performance landscape. For instance, OpenAI moderator achieves a perfect precision score of 1.0000, but with significantly lower scores across other metrics, indicating potential issues balancing precision and recall. AzureAPI, while showing higher performance in terms of ACC compared to OpenAI moderator, exhibits lower precision and recall, highlighting trade-offs between different evaluation criteria. Overall, Table~\ref{tab:aggregateResults} provides a comprehensive overview of the classification results across various models, shedding light on their strengths and weaknesses in different operational contexts.

It should be noted that comparing aggregate metrics can lack rigour as it is not disclosed which datasets have been used for training both open-weight models and closed source ones. Best practice would be to evaluate competitors model on all the datasets present in the literature. Nevertheless, such large scale benchmarking is out of scope for this work since we aim to evaluate the learning capability of naive tabular classifiers on guardrail tasks.  For this reason, hereinafter, we present the dataset-specific model result in terms of TP rate and FP rate for jailbreak and benign datasets respectively.

Figure~\ref{fig:histogramTP}, depicts the true positive (TP) rate for each jailbreak dataset across different models. The first insight observed is that, notably, our tabular models demonstrate varying levels of performance across different datasets, with MoJE exhibiting the most stable TP rate across all datasets. Conversely, the open-weight models, ProtectAI and Llama-Guard, exhibit mixed performance, demonstrating relatively low TP rates compared to the tabular models, particularly evident in datasets such as ``harmful behaviors'' and ``gandalf ignore instructions''. Similarly, the closed-source models, OpenAI moderator and Azure AI Content Safety, show distinct patterns of performance across the datasets, with Azure AI Content Safety demonstrating higher TP rates in ``gandalf ignore instructions" and ``jailbreak prompts", while OpenAI moderator performs relatively poor across all datasets. Overall, the figure provides valuable insights into the TP rates of different models across various jailbreak datasets, highlighting the performance variations and potential areas for improvement in jailbreak detection systems.

On the other hand, Figure~\ref{fig:histogramFP} illustrates the false positive (FP) rate for each benign dataset. Notably, the FP rates vary among the different datasets and models but generally remain very low. For instance, LR exhibits relatively high FP rates across all datasets compared to other models, with notable spikes observed in ``puffin" and ``alpaca'' datasets. Conversely, MoJE demonstrates low FP rates across all datasets, showcasing its superior performance in mitigating false positives. Open-weight models, such as ProtectAI and Llama-Guard, show mixed performance, with ProtectAI exhibiting higher FP rates in the ``puffin'' dataset compared to Llama-Guard. Moreover, closed-source models, OpenAI moderator and Azure AI Content Safety, exhibit distinct patterns of FP rates across the datasets, further emphasizing the variability in model performance. Overall, the figure provides valuable insights into the FP rates of different models across various benign datasets, highlighting their strengths and weaknesses in different operational contexts.

\noindent\textsc{\bfseries Summary.} \textit{\ul{In conclusion, MoJE demonstrates superior performance in balancing precision and recall, highlighting its effectiveness in guardrail tasks for LLMs. The findings underscore the importance of careful model selection and dataset-specific evaluation for robust guardrail systems.}}
\section{Ablation}\label{sec:abl6}

\subsection{Tokenizer and N\_gram Dimension}

Table \ref{tab:tokenizer_ngram} presents the performance comparison of the MoJE architecture across different tokenizers and feature engineering functions $h(\cdot)$, which control and determine the size, $m$, of the extracted input features. Performance is evaluated using various metrics including AUC, accuracy (ACC), $F_{\beta}$ score with $\beta=0.5$, F1 score, recall, and precision. Notably, the choice of tokenizer and feature engineering function significantly influences the curse of dimensionality ($m$), with lower values indicating better performance. For instance, using character-based tokenization with uni-gram features consistently yields the lowest $m$ values across all models, indicating effective dimensionality reduction. Conversely, employing bi-gram features with character-based tokenization results in higher $m$ values, resulting in increased dimensionality and potential overfitting.

Regarding model performance metrics, certain combinations of tokenization and feature engineering demonstrate superior performance across different evaluation metrics. For example, using the BERT tokenizer with uni-gram features yields competitive results across various metrics, achieving the highest AUC and $F_{\beta}$ scores in some cases. Additionally, the combination of uni-gram features with the (uni+bi)-gram tokenizer consistently delivers strong performance across different metrics, highlighting the effectiveness of combining multiple tokenization strategies.

However, it's important to note that the optimal tokenizer-feature engineering combinations vary depending on the evaluation metric. For instance, while certain configurations excel in terms of AUC and $F_{\beta}$ score, they may exhibit lower accuracy or precision. Therefore, practitioners should carefully select the tokenizer and feature engineering function based on the specific requirements of their application and the importance assigned to different evaluation metrics. Overall, the table provides valuable insights into the impact of tokenization and feature engineering on model performance, facilitating informed decision-making in the design and optimization of machine learning architectures.
\begin{table*}[!t]\centering
\begin{tabular}{lrrrrrrrrr}\toprule
Tokenizer &$h(\cdot)$ & $m\downarrow$ &AUC$\uparrow$ &ACC$\uparrow$ &$F_\beta\uparrow$ &F1$\uparrow$ &Recall$\uparrow$ &Precision$\uparrow$ \\\midrule
\multirow{4}{*}{Char} &uni-gram &\textbf{370} &0.9811 &0.9796 &0.8131 &0.7257 &0.6154 &0.8841 \\
&TF-IDF uni-gram &\textbf{370} &0.9808 &0.9803 &0.8346 &0.7281 &0.6004 &0.9249 \\
&bi-gram &\underline{4966} &\underline{0.9978} &0.9933 &0.9488 &0.9202 &0.8762 &0.9689 \\
&(uni+bi)-gram &5336 &\textbf{0.9984} &0.9932 &0.9473 &0.9193 &0.8762 &0.9669 \\ \midrule
\multirow{5}{*}{Word} &uni-gram &30298 &0.9946 &\underline{0.9944} &0.9529 &0.9341 &0.9043 &0.9659 \\
&TF-IDF uni-gram &30298 &0.9964 &0.9933 &0.9467 &0.9205 &0.8799 &0.9650 \\
&bi-gram &271703 &0.9723 &0.9897 &0.9235 &0.8718 &0.7974 &0.9615 \\
&(uni+bi)-gram &302001 &0.9954 &0.9938 &0.9465 &0.9273 &0.8968 &0.9598 \\
\midrule
\multirow{4}{*}{BERT} &uni-gram &16964 &0.9926 &0.9933 &0.9328 &0.9226 &\underline{0.9062} &0.9397 \\
&TF-IDF uni-gram &16964 &0.9960 &\underline{0.9944} &0.9529 &0.9341 &0.9043 &0.9659 \\
&bi-gram &283091 &0.9862 &0.9913 &0.9376 &0.8938 &0.8293 &0.9693 \\
&(uni+bi)-gram &300055 &0.9970 &\textbf{0.9946} &\textbf{0.9564} &\underline{0.9370} &0.9062 &0.9699 \\\midrule
\multirow{4}{*}{GPT2} &uni-gram &22493 &0.9943 &0.9937 &0.9450 &0.9264 &0.8968 &0.9579 \\
&TF-IDF uni-gram &22493 &0.9971 &0.9932 &0.9504 &0.9188 &0.8705 &0.9727 \\
&bi-gram &292105 &0.9799 &0.9900 &0.9321 &0.8749 &0.7936 &\underline{0.9747} \\
&(uni+bi)-gram &314598 &0.9899 &0.9926 &0.9412 &0.9111 &0.8649 &0.9624 \\\midrule
\multirow{4}{*}{Llama2} &uni-gram &10354 &0.9952 &0.9938 &0.9445 &0.9275 &0.9006 &0.9562 \\
&TF-IDF uni-gram &10354 &0.9972 &0.9938 &0.9465 &0.9273 &0.8968 &0.9598 \\
&bi-gram &257156 &0.9785 &0.9906 &0.9371 &0.8827 &0.8049 &\textbf{0.9772} \\
&(uni+bi)-gram &267510 &0.9956 &\textbf{0.9946} &\underline{0.9533} &\textbf{0.9373} &\textbf{0.9118} &0.9643 \\
\bottomrule
\end{tabular}\caption{Difference of performance of the MoJE architecture using different tokenizers and different feature engineering function (i.e., $h(\cdot)$. We can see how Tokenization and $h(\cdot)$ can results in different features extracted dimension (i.e., $m$) which it means the lower the feature extracted the most efficient the training and inference. Best results are in bold and second-best underlined.}\label{tab:tokenizer_ngram}
\end{table*}

\begin{figure*}[!ht]
    \centering
    \captionsetup[subfigure]{font=scriptsize,labelfont=scriptsize}
\subfloat[]{
    \begin{tikzpicture}
    \begin{scope}[scale=0.63, transform shape]
        \begin{axis}[
        scaled x ticks=true,
            xmajorgrids,
            ymajorgrids,
            xlabel={$\mathbf{m}$},
            ylabel={$\mathbf{F_{\beta}}$},
            ymin=0.80,
            ymax=0.99,
            xmin=0,
            enlarge x limits=0.15,
        ]
        \addplot[green,draw=black,mark=triangle*, mark size=3pt] coordinates {(370, 0.8131)};
        \addplot[green,draw=black,mark=diamond*,mark size=3pt] coordinates {(370, 0.8346)};
        \addplot[green,draw=black,mark=pentagon*,mark size=3pt] coordinates {(4966, 0.9488)};
        \addplot[green,draw=black,mark=square*, mark size=3pt] coordinates {(5336, 0.9473)};

        \end{axis}
        \end{scope}
    \end{tikzpicture}}\label{fig:feature1} \quad
\subfloat[]{
    \begin{tikzpicture}
    \begin{scope}[scale=0.63, transform shape]
        \begin{axis}[
        scaled x ticks=true,
            xmajorgrids,
            ymajorgrids,
            xlabel={$\mathbf{m}$},
            ylabel={$\mathbf{F_{\beta}}$},
            ymin=0.80,
            ymax=0.99,
            xmin=10000,
            enlarge x limits=0.15,
        ]

        \addplot[red,draw=black,mark=triangle*, mark size=3pt] coordinates {(30298, 0.9529)};
        \addplot[red,draw=black,mark=diamond*, mark size=3pt] coordinates {(30298, 0.94671)};
        
        \addplot[blue,draw=black,mark=triangle*, mark size=3pt] coordinates {(16964, 0.9328)};
        \addplot[blue,draw=black,mark=diamond*,mark size=3pt] coordinates {(16964, 0.9529)};

        \addplot[orange,draw=black,mark=triangle*, mark size=3pt] coordinates {(22493, 0.9450)};
        \addplot[orange,draw=black,mark=diamond*,mark size=3pt] coordinates {(22493, 0.9504)};

        \addplot[purple,draw=black,mark=triangle*, mark size=3pt] coordinates {(10354, 0.9445)};
        \addplot[purple,draw=black,mark=diamond*,mark size=3pt] coordinates {(10354, 0.9465)};

        \end{axis}
        \end{scope}
    \end{tikzpicture}}\label{fig:feature2} \quad
\subfloat[]{
    \begin{tikzpicture}
    \begin{scope}[scale=0.63, transform shape]

        \begin{axis}[
        scaled x ticks=true,
            xmajorgrids,
            ymajorgrids,
            xlabel={$\mathbf{m}$},
            ylabel={$\mathbf{F_{\beta}}$},
            ymin=0.80,
            ymax=0.99,
            xmin=255000,
            enlarge x limits=0.15,
        ]


        \addplot[red,draw=black,mark=pentagon*,mark size=3pt] coordinates {(271703, 0.923512)};
        \addplot[red,draw=black,mark=square*,mark size=3pt] coordinates {(302001, 0.946535)};
        

        \addplot[blue,draw=black,mark=pentagon*,mark size=3pt] coordinates {(283091, 0.9376)};
        \addplot[blue,draw=black,mark=square*, mark size=3pt] coordinates {(300055, 0.9564)};

        \addplot[orange,draw=black,mark=pentagon*,mark size=3pt] coordinates {(292105, 0.9321)};
        \addplot[orange,draw=black,mark=square*, mark size=3pt] coordinates {(314598, 0.9412)};

        \addplot[purple,draw=black,mark=pentagon*,mark size=3pt] coordinates {(257156, 0.9371)};
        \addplot[purple,draw=black,mark=square*, mark size=3pt] coordinates {(267510, 0.9533)};

        \end{axis}
    \end{scope}
    \end{tikzpicture}
    } \label{fig:feature3} \quad

\vspace{0.3em}
\begin{tikzpicture}
        \begin{customlegend}[legend columns=4,
        legend style={
        at={(0, 0.2)},
       anchor=north,
       inner sep=3pt,
       draw=none,
       style={column sep=0.15cm},
       name=leg1},
        legend entries={\footnotesize uni-gram, \footnotesize TF-IDF uni-gram, \footnotesize bi-gram, \footnotesize (uni+bi)-gram}]
        \addlegendimage{only marks, white, draw=black, mark=triangle*, mark size=3pt}
        \addlegendimage{only marks, white, draw=black, mark=diamond*, mark size=3pt}
        \addlegendimage{only marks, white, draw=black, mark=pentagon*, mark size=3pt}
        \addlegendimage{only marks, white, draw=black, mark=square*, mark size=3pt}
        \end{customlegend}
        \begin{customlegend}[legend columns=3,
        legend style={
        at={(0, -0.33)},
       anchor=north,
       inner sep=3pt,
       draw=none,
       style={column sep=0.15cm},
       name=leg2},
        legend entries={\footnotesize Char, \footnotesize Word, \footnotesize BERT,  \footnotesize GPT2, \footnotesize Llama2}]
        \addlegendimage{area legend, fill=green}
        \addlegendimage{area legend, fill=red}
        \addlegendimage{area legend, fill=blue}
        \addlegendimage{area legend, fill=orange}
        \addlegendimage{area legend, fill=purple}
        \end{customlegend}
        \node [fit=(leg1)(leg2),draw] {};
    \end{tikzpicture}
    \caption{Relation between number of feature extracted (i.e., $m$) and $F_{\beta}$ based on the different Tokenizer (i.e., \textit{Char}, \textit{Word}, \textit{BERT}, \textit{GPT2}, and \textit{Llama2}) and features extraction functions (i.e., \textit{uni-gram}, \textit{TF-IDF uni-gram}, \textit{bi-gram}, \textit{uni+bi-gram}). The figure is divided into three sub-figures due to the sparsity of the results with respect to the x-axis.}
    \label{fig:plot_ngram_points}
\end{figure*}
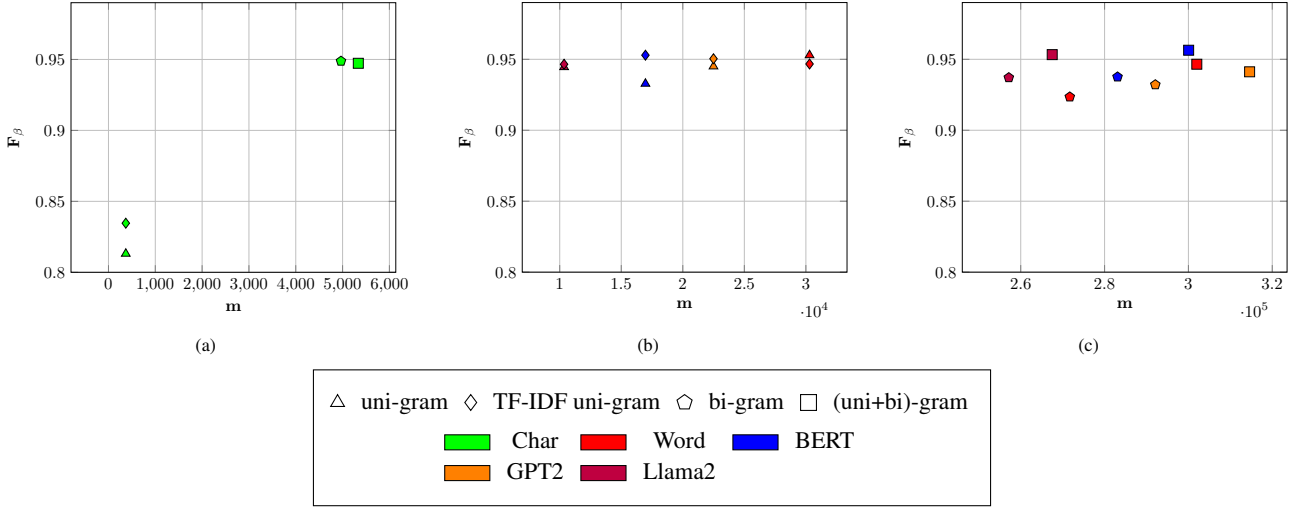

Figure \ref{fig:plot_ngram_points} illustrates the relationship between the number of features extracted ($m$) and the $F_{\beta}$ score across different tokenizers (i.e., Char, Word, BERT\footnote{\url{https://huggingface.co/google-bert/bert-base-uncased}}, GPT2\footnote{\url{https://huggingface.co/openai-community/gpt2}}, and Llama2\footnote{\url{https://huggingface.co/meta-llama/Llama-2-7b-hf}}) and feature extraction functions (i.e., uni-gram, TF-IDF uni-gram, bi-gram, uni+bi-gram). Each subplot (a), (b), and (c) corresponds to a specific tokenizer, showcasing how the choice of feature extraction function influences the number of dimensions ($m$) and subsequently impacts the $F_{\beta}$ score.

In each subplot, distinct markers represent different feature extraction functions (i.e., $h(\cdot)$), with variations in color indicating the different tokenizers used. The legend provides a clear mapping between marker types and feature extraction functions, facilitating easy interpretation of the results. Notably, the figure highlights the trade-off between the number of features extracted and the $F_{\beta}$ score, with certain combinations of tokenizers and feature extraction functions demonstrating more favorable trade-offs.

The visualization aids in identifying optimal configurations that balance feature dimensionality reduction with model performance. By analyzing the trends across different tokenizers and feature extraction functions, practitioners can make informed decisions when selecting the appropriate combination for their specific machine learning tasks. Overall, the figure provides valuable insights into the impact of tokenization and feature extraction on model performance, contributing to the optimization of machine learning architectures.

\noindent\textsc{\bfseries Summary.} \textit{\ul{The comparison table underscores the importance of selecting appropriate tokenization and feature engineering strategies in optimizing MoJE architecture performance. Similarly, the visualization highlights the trade-offs between feature dimensionality and model performance, providing valuable insights for practitioners to make informed decisions in designing machine learning architectures.}}

\subsection{Mutual Information Theorem for Feature Selection}
In the previous section, we have noticed how the dimensions of a corpus, the chosen tokenization method, and the feature extraction approach can all contribute to the curse of dimensionality. Classical machine learning techniques such as feature selection and dimensionality reduction offer viable strategies to address this challenge.
In our work, we decided to employ the Mutual Information Gain theorem as feature selection strategy~\cite{DBLP:conf/ijcnn/BerahaMPTR19}. The aforementioned theorem, initially proposed by~\cite{quinlan1986induction}, assesses the discrepancy between the entropy of class distribution and the conditional entropy given a specific feature.

Specifically, the mutual information (MI) between two random variables $X$ (i.e., our extracted features) and $Y$ (i.e., our target class) is defined as:
\begin{equation}
I(X ; Y) :=H(Y)-H(Y \mid X),
\end{equation}

where the entropy $H(X)$ of a random variable $X$, having $p$ as probability density function, is a measure of uncertainty:
$$
H(X):=\mathbb{E}_X[-\log (p(X))]=-\int p(x) \log p(x) d x .
$$

Intuitively, the MI between $X$ and $Y$ represents the reduction in the uncertainty of $Y$ after observing $X$ (and vice-versa).

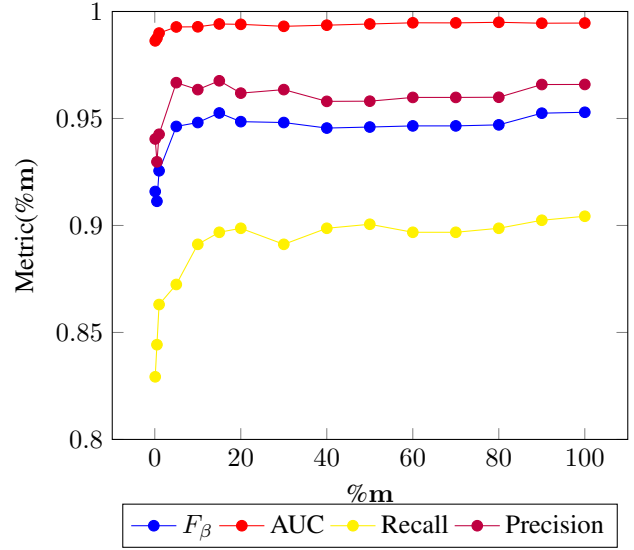
\begin{figure}[htbp]
    \centering
    \begin{tikzpicture}
        \begin{axis}[
            xlabel={\%$\mathbf{m}$},
            ylabel={Metric(\%$\mathbf{m}$)},
            ymin=0.8,
            ymax=1,
            enlarge x limits=0.1,
            legend style={at={(0.5,-0.15)},
                anchor=north,legend columns=-1},
        ]
        
        \addplot[blue,mark=*] coordinates {
            (0.1, 0.915872)
            (0.5, 0.9113)
            (1, 0.925553)
            (5, 0.946276)
            (10, 0.948104)
            (15, 0.952571)
            (20, 0.948515)
            (30, 0.948104)
            (40, 0.945519)
            (50, 0.945999)
            (60, 0.946535)
            (70, 0.946535)
            (80, 0.947015)
            (90, 0.952475)
            (100, 0.9529)

        };
        \addplot[red,mark=*] coordinates {
            (0.1, 0.986272)
            (0.5, 0.987492)
            (1, 0.98992)
            (5, 0.992772)
            (10, 0.992844)
            (15, 0.994165)
            (20, 0.993994)
            (30, 0.993074)
            (40, 0.993599)
            (50, 0.994155)
            (60, 0.994721)
            (70, 0.994677)
            (80, 0.994947)
            (90, 0.994519)
            (100, 0.9946)
        };
        \addplot[yellow,mark=*] coordinates {
            (0.1, 0.829268)
            (0.5, 0.844278)
            (1, 0.863039)
            (5, 0.87242)
            (10, 0.891182)
            (15, 0.896811)
            (20, 0.898687)
            (30, 0.891182)
            (40, 0.898687)
            (50, 0.900563)
            (60, 0.896811)
            (70, 0.896811)
            (80, 0.898687)
            (90, 0.902439)
            (100, 0.9043)

        };
        \addplot[purple,mark=*] coordinates {
            (0.1, 0.940426)
            (0.5, 0.929752)
            (1, 0.942623)
            (5, 0.966736)
            (10, 0.963489)
            (15, 0.967611)
            (20, 0.961847)
            (30, 0.963489)
            (40, 0.958)
            (50, 0.958084)
            (60, 0.959839)
            (70, 0.959839)
            (80, 0.95992)
            (90, 0.965863)
            (100, 0.9659)
        };
        \legend{$F_{\beta}$, AUC, Recall, Precision}
        \end{axis}
    \end{tikzpicture}
    \caption{Effect on AUC, $F_\beta$, Recall, and Precision with different percentage of feature selected (i.e., \%$\mathbf{m}$) ranked according to the Mutual Information Theorem. The base model refer to \textit{Word} as tokenizer and uni-gram as $h(\cdot)$ where 100\% of $m$ correspond to 30298 features (see Table~\ref{tab:tokenizer_ngram}). }
    \label{fig:plot_mi_curve}
\end{figure}
Figure~\ref{fig:plot_mi_curve} illustrates the impact of feature selection, based on the MI Theorem, on several performance metrics, including AUC, $F_{\beta}$, Recall, and Precision. Each curve represents the variation in these metrics as the percentage of selected features, denoted as percentage of $\mathbf{m}$, changes. Four distinct colored lines depict the performance trends for each metric, providing insights into how different percentages of feature selections influence model performance. The x-axis represents the percentage of features selected, while the y-axis shows the corresponding values of the performance metrics.

As the percentage of $\mathbf{m}$ increases, the performance metrics exhibit diverse behaviors. Notably, for AUC and $F_{\beta}$, there is a consistent improvement with an increasing percentage of features selected, indicating that a higher number of informative features contributes to better model performance in terms of overall classification accuracy and balance between precision and recall. Conversely, Recall and Precision show more nuanced patterns, with Recall generally increasing as more features are selected, while Precision may exhibit fluctuations or reach a plateau after a certain percentage of $\mathbf{m}$. These observations suggest that while increasing feature selection can enhance certain aspects of model performance, there may be diminishing returns or trade-offs to consider, particularly in terms of Precision.

Overall, the figure provides valuable insights into the relationship between feature selection and model performance, highlighting the importance of optimizing the percentage of $\mathbf{m}$ to achieve the desired balance between various performance metrics. Additionally, the results underscore the effectiveness of utilizing the MI Theorem as a feature selection strategy, demonstrating its ability to improve the discriminative power of the model while maintaining a balance between different evaluation criteria.

\noindent\textsc{\bfseries Summary.} \textit{\ul{
We demonstrate how feature selection using the Mutual Information Theorem impacts key performance metrics, revealing nuanced behaviors in AUC, $F_{\beta}$, Recall, and Precision as the percentage of selected features changes, highlighting the need for optimizing feature selection to achieve a balanced model performance.}}


\subsection{Out of Distribution Data and MoJE Modularity}


With new LLM attacks being developed, updating guardrails can require significant time and computational power specifically for LLM based guardrails (e.g., Llama-Guard). Furthermore, training this model remains challenging since engineers should decide the right compromise between time and performance due to poor hyperparameter exploration.
Our model overcomes all these challenges through its inherent modular architecture.

Let's consider a case in which we need to integrate new datasets that are out of distribution (OOD) with respect to the one used for training due to poor performance. Our architecture allows us to train a new classifier $f_{\theta}^{l+1}(\cdot)$ and add it into the mix of MoJE as the $l+1$-th classifier. As an example, we will consider two jailbreak datasets (i.e., ``\textit{aart}'' and ``\textit{attaq}''), one benign dataset (i.e., ``\textit{boolq}''), and a mix of extreme cases of jailbreak and benign prompts (i.e., ``\textit{xstest}''). Results are reported in Table~\ref{tab:modularity}.\footnote{We want to highlight that what we consider OOD data for our model may not be the same for open-weight (i.e., ProtectAI and Llama-Gard) and close source models (i.e., openAI moderator and Azure AI Content Safety API) as we do not have access to their training data or training pipeline.}

\begin{table}[!t]\centering
\small
\begin{tabular}{lrrrrrr}\toprule
&\textbf{aart} &\textbf{attaq} &\multicolumn{2}{c}{\textbf{xstest}} &\textbf{boolq} \\
\midrule
$f_{\theta}(X)$&TPr$\uparrow$ &TPr$\uparrow$ &TPr$\uparrow$ &FPr$\downarrow$ &FPr$\downarrow$ \\\midrule
$\mathrm{MoJE}^{l}$ &0.0713 &0.0000 &0.0000 &\textbf{0.0000} &\textbf{0.0000} \\
$\mathrm{MoJE}^{l+3}$ &\textbf{0.9395} &0.6044 &0.4727 &0.0857 &0.0004 \\\midrule
ProtectAI &0.0100 &0.0000 &0.0000 &0.0000 &0.0100 \\
Llama-G. &0.8400 &\textbf{0.8462} &\textbf{0.7818} &0.0857 &\textbf{0.0000} \\\midrule
openAI m. &0.1600 &0.3956 &0.3273 &0.0286 &\textbf{0.0000} \\
AzureAPI &0.0000 &0.0000 &0.0000 &\textbf{0.0000} &\textbf{0.0000} \\
\bottomrule
\end{tabular}\caption{Results on new out of distribution dataset of the presented model $f_\Theta^{MoJE}(\cdot)$ with $l$=4 as number of jailbreak expert and the same with new experts for each jailbreak dataset. Note that \textit{boolq} is not a jailbreak so we are not training an expert with respect to this dataset. Best results are in bold font.}\label{tab:modularity}
\end{table}

Table \ref{tab:modularity} presents the performance of the MoJE model with four jailbreak experts ($l=4$) and with additional experts trained for each new jailbreak dataset, addressing the challenge of integrating OOD data. The table showcases the true positive rates (TPr) and false positive rates (FPr) for various models, including MoJE, ProtectAI, LlamaGuard, OpenAI moderator, and Azure API, across different datasets. Notably, MoJE with additional experts ($l+3$) demonstrates substantial improvements in TPr for \textit{aart} and \textit{attaq} datasets, highlighting the model's adaptability to new OOD data, while maintaining low FPr rates across all datasets.

Nevertheless, considering the dataset \textit{xstest}, which can be considered difficult even for human moderators, our tabular classifier and architecture demonstrates it's linguistic limitation. Only an LLM-based guardrail such as Llama-Guard, with its context learning capability, can partially mitigate certain jailbreak prompt attacks.

\noindent\textsc{\bfseries Summary.} \textit{\ul{
Our model, MoJE, effectively addresses the challenges of easily integrating new classifiers for out-of-distribution datasets. MoJE achieves substantial improvements in true positive rates while maintaining low false positive rates across various new OOD datasets. However, in cases like ``xstest'', where linguistic complexities arise, our tabular classifier shows limitations, highlighting the need for LLM-based guardrails like Llama-Guard for better mitigation of certain jailbreak prompt attacks.}}

\section{Conclusion}\label{sec:concl7}
In conclusion, MoJE demonstrates superior performance in balancing precision and recall, underscoring its effectiveness in guardrail tasks for Large Language Models (LLMs) while maintaining a simple and low-computation resource architecture. The results highlight the significance of careful model selection and dataset-specific evaluation for robust guardrail systems. Through ablation studies, we underscore the importance of selecting appropriate tokenization and feature engineering strategies, optimizing feature selection, and integrating new classifiers for out-of-distribution datasets. The findings reveal nuanced behaviors in performance metrics and emphasize the need for optimizing feature selection to achieve balanced model performance specifically in case of curse of dimensionality. While MoJE effectively addresses various challenges, such as computational efficiency and adaptability to new datasets, it shows limitations in handling complex linguistic prompts like ``\textit{xstest}'', indicating the necessity for LLM-based guardrails like Llama-Guard for better mitigation of certain jailbreak prompt attacks.

Future work will focus on further enhancing the adaptability of MoJE by exploring other low-weight language model architectures, new feature engineering techniques, or linguistic data augmentation techniques. Furthermore, a hybrid approach combining statistical methods with deep learning could be a possible new research direction.

\bibliography{main}

\end{document}